\RequirePackage{rotating}
\documentclass[12pt]{iopart}
\usepackage{makeidx}
\usepackage{graphicx}
\usepackage{amssymb,amsfonts}
\usepackage{verbatim}
\usepackage{nameref,hyperref} 
\usepackage{makecell}
\usepackage{color,xcolor}
\usepackage{cite}
\usepackage{setspace}
\usepackage{bm}
\def\erfc{\textrm{erfc}}

\begin{document}

\ifx\aligned\undefined

\makeatletter
\def\aligned{{\ifnum0=`}\fi\vcenter\bgroup\let\\\cr
\halign\bgroup&\hfil$\displaystyle{}##{}$&$\displaystyle{}##{}$\hfil\cr}
\def\endaligned{\crcr\egroup\egroup\ifnum0=`{\fi}}
\def\align{\par
\bigskip
{\ifnum0=`}\fi
\let\\\cr
\halign to \textwidth\bgroup
\refstepcounter{equation}%
\global\let\@alignlab\@currentlabel
\vrule \@height \dimexpr\ht\strutbox+3pt\relax
      \@depth  \dimexpr\dp\strutbox+1pt\relax
      \@width \z@
\hbox to \textwidth{\hfill(\theequation)}\kern-\textwidth
\tabskip\fill
\hfil$\displaystyle{}##{}$&%
\let\@currentlabel\@alignlab$\displaystyle{}##{}$\hfil&%
\let\@currentlabel\@alignlab\hfil$\displaystyle{}##{}$&%
\let\@currentlabel\@alignlab$\displaystyle{}##{}$\hfil\cr}
\def\endalign{\crcr\egroup\ifnum0=`{\fi}\par\bigskip}
\newcounter{parentequation}
\newenvironment{subequations}{%
  \refstepcounter{equation}%
  \edef\theparentequation{\theequation}%
  \setcounter{parentequation}{\value{equation}}%
  \setcounter{equation}{0}%
  \def\theequation{\theparentequation\alph{equation}}%
  \ignorespaces
}{%
  \setcounter{equation}{\value{parentequation}}%
  \ignorespacesafterend
}
\def\cases{{\ifnum0=`}\fi\left\{\array{lll}}
\def\endcases{\endarray\right.\ifnum0=`{\fi}}
\makeatother
\fi

\title{Non-Gaussian diffusion of mixed origins}
\author{Yann Lanoisel\'ee}
  \ead{y.lanoiselee@bham.ac.uk}

\address{Institute of Metabolism and Systems Research, University of Birmingham, Birmingham, UK \\ and}%
\address{Centre of Membrane Proteins and Receptors (COMPARE), Universities of Birmingham and Nottingham, UK \\ and}
\address{Laboratoire de Physique de la Mati\`{e}re Condens\'{e}e (UMR 7643), \\ CNRS -- Ecole Polytechnique, IP Paris, 91128 Palaiseau, France}
\author{Denis S. Grebenkov}
\ead{denis.grebenkov@polytechnique.edu}
\address{Laboratoire de Physique de la Mati\`ere Condens\'ee, \\ CNRS -- Ecole Polytechnique, IP Paris, 91128 Palaiseau, France}

\date{\today}

\begin{abstract}
The properties of diffusion processes are drastically affected by heterogeneities of the medium that can induce non-Gaussian behavior of the propagator in contrast with the idealized realm of Brownian motion. In this paper we analyze the diffusion propagator when distinct origins of heterogeneity (e.g. time-fractional diffusion, diffusing diffusivity, distributed diffusivity across a population) are combined. These combinations allow one to describe new classes of strongly heterogeneous processes relevant to biology. Based on a combined subordination technique, we obtain the exact propagator for different instructive examples. This approach is then used to calculate analytically the first-passage time statistics (on half-real line and in any bounded domain) for a particle undergoing non-Gaussian diffusion of mixed origins.
\end{abstract}


\date{\today}
\maketitle
\doublespacing
\section{Introduction}
The description of diffusion processes in heterogeneous media is fundamental for characterizing physical and biological diffusion-controlled systems.
In the last decade, many experiments in complex media have shown evidences that heterogeneities have a drastic effect onto the diffusive properties of particles. In particular, the distribution of displacements of a particle can significantly deviate from the conventional Gaussian distribution \cite{Orpe2007,Gollub1991,Stuhrmann2012,Toyota2011,Bertrand2012,Moschakis2012,Chaudhuri2007,Grady2017,Wang2009,Wang2012,Rouyer2000,He2016,Sarnanta2016}. The non-Gaussian character, observed at short times, can either disappear at long times or not.
In the former case, the particle can be assumed to explore different diffusivities along its path that induce non-Gaussian displacements. In such theoretical models, when the diffusivity is fluctuating around its mean \cite{Chubynsky2014, Chechkin2017, Tyagi2017, Jain2016, Jain2017b,Lanoiselee2018a, Sposini2018}, the Gaussian regime is recovered when the trajectory duration is much longer than the mean-reverting time of diffusivity \cite{Tyagi2018,Jain2018}. 
In another group of models, known as annealed transit time model (ATTM), the particle experiences a medium where the diffusion coefficient does not evolve continuously but changes at random times \cite{Massignan2014}. 

When diffusion remains non-Gaussian even at long times, several explanations can be at play.
When the distribution of periods with a constant diffusivity in the ATTM model has a heavy tail with infinite mean, the particle never explores the whole distribution of diffusivities. Therefore the dynamics exhibits weak ergodicity breaking \cite{Massignan2014,Manzo2015}. Another model that displays both non-Gaussian distribution and weak ergodicity breaking is the time-fractional diffusion when the particle dynamics is so slow that it never explores the whole phase space. The non-Gaussian behavior can also originate from inter-trajectory fluctuations when trajectories with different diffusion coefficients are used to build up an empirical probability density \cite{Beck2003,Beck2005,Vitali2018}. This situation can typically happen in cell biology when inter-cell fluctuations are unavoidable. Another occurrence of non-Gaussian distributions arises when a Gaussian process is superimposed with additional non-Gaussian noise. \textit{A priori}, all these mechanisms are not mutually exclusive and can combine to produce very complicated types of motion. 

In this article, we investigate the probability density function of particle displacements subject to non-Gaussian diffusion of mixed origins when several mechanisms contribute. Previously, mixed diffusion processes has already been discussed for the particular cases of Continuous Time Random Walk on a fractal structure \cite{Meroz2010} and subordinated fractional Brownian motion \cite{Thiel2014,Tabei2013}. We will take advantage of the fact that many non-Gaussian diffusion processes can be mathematically described via subordination \cite{Bochner1949,Fogedby1994,Weron2008,Meerschaert2013,Kumar2018a,Kumar2018b}. Then we successively apply the subordination technique in order to describe the combined effect of different mechanisms causing heterogeneous non-Gaussian diffusion. The path properties of the particle remain the same as in the case of Brownian motion but the ``speed'' at which the particle travels along the path is disorder-driven. An equivalent way of interpreting the subordination is that the integrated diffusivity is not deterministic but randomly increasing.

Sec. \ref{sec:int_diff_representation}, summarizes in a systematic way several formerly known physical mechanisms leading to non-Gaussian diffusion in terms of subordination integrals and describes the particular forms of the probability density function of their integrated diffusivity. In Sec. \ref{sec:CTR_integrated_diffusivity}, we introduce a model of integrated diffusivity which can either converge to normal diffusion at long time or not. We also discuss its relation to other random diffusivity models. In Sec. \ref{sec:multiple_subordination}, using the properties of subordination, we obtain the probability density function of particle displacements, subject to several heterogeneity origins. In Sec. \ref{sec:first_passage}, we address an important topic of first-passage time statistics that have been studied for diffusing diffusivity \cite{Jain2016b,Lanoiselee2018b,Sposini2018b}, Brownian motion subordinated by a L\'{e}vy process \cite{Hurd2009} and general random diffusivity models \cite{Lanoiselee2018b,Grebenkov2018} but not for their combination. Based on our description, we calculate, for these mixed models, the first-passage time statistics in a bounded domain and on the real half-axis.
\section{Integrated-diffusivity zoology}
\label{sec:int_diff_representation}
In this section we make a non-exhaustive inventory of models of non-Gaussian diffusion that can be described via the subordination approach. For the sake of clarity, we focus on one-dimensional diffusion here but most statements remain valid in any dimension (see, e.g., Sec. \ref{sec:first_passage_bounded_domain}). In order to compute the probability density function $P(x,t)$ of being at $x$ at time $t$ starting from the origin at time $0$, one specifies the probability density $p(x,u)$ of a displacement $x$ with integrated diffusivity $u$ and integrate it to the probability density function $R(u,t)$ of having $u$ in a time $t$ such that
\begin{equation}\label{eq:Brownian_subordination_integral}
    P(x,t)=\int\limits_0^\infty p(x,u)R(u,t)du.
\end{equation}
While the integrated diffusivity $u$ has units of squared length, it is sometimes convenient to rescale it by diffusivity to get units of time. Whatever the convention is used, $u$ will always appear as a dimensionless ratio $u/\mu_u$, where $\mu_u$ is the scale of integrated diffusivity. Therefore, in the present form, $u$ can be interpreted equivalently as an integrated diffusivity, as an operational time, or as a superposition of independent processes. We will use this property in Sec. \ref{sec:multiple_subordination}, to combine different models of disorder.

In the case of one-dimensional Brownian motion, the diffusion coefficient is constant so the integrated diffusivity is deterministic $u=\mu_u t/\mu_t$, where $\mu_u$ and $\mu_t$ are respectively the scales of integrated diffusivity and of physical time, from which we have
\begin{equation}\label{eq:Ru_BM}
    R(u,t)=\delta(u-\mu_u t/\mu_t),
\end{equation}
and the distribution of displacement reads
\begin{equation}\label{eq:gaussian_disp_distrib}
p(x,u)=\frac{1}{2\sqrt{\pi\mu_x^2 u/\mu_u}}\exp\left(-\frac{x^2}{4\mu_x^2 u/\mu_u}\right),
\end{equation}
with the length-scale $\mu_x$. After integration in Eq. (\ref{eq:Brownian_subordination_integral}), we retrieve the Gaussian propagator
\begin{equation}\label{eq:BM_distrib}
P(x,t)=\frac{1}{2\sqrt{\pi Dt}}\exp\left(-\frac{x^2}{4Dt}\right),
\end{equation}
where the diffusion coefficient $D=\mu_x^2/\mu_t$ is the ratio of the squared length-scale $\mu_x^2$ to the time-scale $\mu_t$. Note that the scale associated to $u$ disappears after integration because we chose the same scale $\mu_u$ in Eqs. (\ref{eq:Ru_BM},\ref{eq:gaussian_disp_distrib}); along the rest of the paper the final propagators
will include only $\mu_x$ and $\mu_t$ due to the consistent choice of intermediate scales.


\subsection{Time-fractional diffusion}
Perhaps the best studied model of annealed disorder leading to non-Gaussian distribution is the one of time-fractional diffusion of order $\beta$. This anomalous diffusion process that displays weak ergodicity breaking can be obtained as the long-time limit of a continuous time random walk in which the waiting times are power-law distributed with an infinite mean. The probability density function of displacements is expressed in terms of the Fox H function by using a Mellin transform of the time-fractional diffusion equation \cite{Wyss1986}. The very same results can equivalently be obtained by subordination of Brownian motion with an inverse L\'evy stable subordinator \cite{Meerschaert2013, Weron2008}. The pathwise representation to time-fractional diffusion equation is 
\begin{equation}
    x(t)=y(S(t)),
\end{equation}
with $y(t)$ the standard Brownian motion (or Wiener process) and $S(t)$ the subordinator
\begin{equation}\label{eq:def_subor_CTRW}
    S(t)=\inf\lbrace \tau>0:U_\beta(\tau)>t\rbrace,
\end{equation}
where $U_\beta(\tau)$ is the stable subordinator with an exponent $0<\beta<1$.
The associated propagator $P(x,t)$  takes the form of a subordination integral
\begin{equation} \label{eq:CTRW_subor}
P(x,t)=\int\limits_0^\infty p(x,\tau)T(\tau,t)d\tau,
\end{equation}
where $p(x,\tau)$ plays the same role as $p(x,u)$ in Eq. (\ref{eq:Brownian_subordination_integral}) (i.e. the probability density of displacements) and $T(\tau,t)$ is the probability density of the integrated operational time 
\begin{equation}\label{eq:integ_optime_def}
    T(\tau,t)=\frac{\partial}{\partial \tau}\mathbb{P}\left\lbrace \int\limits_0^t S(t') dt'<\tau\right\rbrace.
\end{equation}

In the case of time-fractional diffusion, the probability density $T(\tau,t)$ of the inverse stable subordinator is given by \cite{Meerschaert2013}
\begin{equation}
T(\tau,t)=\frac{t}{\beta\tau^{1+1/\beta}}\frac{\mu_\tau^{1/\beta}}{\mu_t}f_\beta\left(\frac{t\mu_\tau^{1/\beta}}{\mu_t\tau^{1/\beta}}\right),
\end{equation}
where $\mu_\tau$ is the time-scale of internal time and $f_\beta(z)$ is a one-sided L\'{e}vy distribution defined by the inverse Laplace transform of $\hat{f_\beta}(s)=\exp\left(-s^\beta\right)$. The exact form of $f_\beta(z)$ is known under different equivalent forms

\begin{equation}
\fl\displaystyle
    f_\beta(z)= \frac{1}{z}\scriptstyle H_{1,1}^{1,0}\left[\displaystyle z^{-\beta}\scriptstyle\middle\vert
\begin{array}{cc}
(0,\beta)\\(0,1) 
\end{array}
\right]\displaystyle
=
\frac{1}{\beta}\scriptstyle H_{1,1}^{1,0}\left[\displaystyle z^{-1} \scriptstyle \middle\vert
\begin{array}{cc}
(1,1)\\(1/\beta,1/\beta) 
\end{array}
\right]\displaystyle
=
\frac{1}{\beta z^2}\scriptstyle H_{1,1}^{1,0}\left[\displaystyle z^{-1}\scriptstyle\middle\vert
\begin{array}{cc}
(-1,1)\\(-1/\beta,1/\beta) 
\end{array}
\right],
\end{equation}
where $H_{1,1}^{1,0}\left[x\middle\vert
\begin{array}{cc}
(a,A)\\(b,B) 
\end{array}
\right]
$ is the Fox H function. The first and second forms can be found in \cite{Mathai2010} and the last in \cite{Barkai2001} and references therein. As it simplifies Mellin transform calculations (see Sec. \ref{sec:SDT_fractional_diffusion}), the first form will be used in the rest of the paper. One gets thus
\begin{equation}\label{eq:CTRW_subordinator}
T(\tau,t)=\displaystyle \frac{1}{\beta \tau }H_{1,1}^{1,0}\left[\frac{\mu_t^\beta\tau}{t^\beta\mu_\tau}\middle\vert
\begin{array}{cc}
(0,\beta)\\(0,1) 
\end{array}
\right].
\end{equation}

\subsection{Diffusing diffusivity}
\label{sec:integrated_diffusing_diffusivity_description}
The subordination concept has also been used to describe the propagator $P(x,t)$ of a diffusion process with time-dependent diffusivity \cite{Chechkin2017,Lanoiselee2018a}. In this case, $P(x,t)$ is calculated by another subordination integral 
\begin{equation} \label{eq:diff_diff_subor}
P(x,t)=\int\limits_0^\infty p(x,u)Q(u,t)du,
\end{equation}
where $Q(u,t)$ is interpreted as the probability density of integrated diffusivity
\begin{equation}\label{eq:integ_diff_def}
Q(u,\tau)=\frac{\partial}{\partial u}\mathbb{P}\left\lbrace \int\limits_0^t D(t') dt'<u\right\rbrace.
\end{equation}
Several models have been explored in which the mean squared displacement is linear but the diffusivity is fluctuating around its mean \cite{Chubynsky2014,Jain2016,Jain2017b,Tyagi2017,Chechkin2017,Lanoiselee2018a}. For instance, diffusivity was modeled as the distance from the origin of an $n-$dimensional Ornstein-Uhlenbeck process \cite{Jain2016,Jain2017b,Tyagi2017,Chechkin2017}, as a Cox-Ingersoll-Ross (CIR) process \cite{Lanoiselee2018a}, as an Ornstein-Uhlenbeck process with right-sided L\'{e}vy noise \cite{Jain2017a}, with stretched exponential distribution of diffusivity \cite{Sposini2018} or as the square of a Gaussian process with a non-monotonously decaying autocorrelation function \cite{Tyagi2018}.
The Langevin equation associated to positive time-dependent diffusivity $D_t$ can be written
\begin{equation}\label{eq:Langevin_D}
dD_t=a_D(D_t)dt+b_D(D_t)dW_t,
\end{equation} 
with some drift $a_D$ and volatility $b_D$. The positivity of $D_t$ can be ensured either by choosing the specific form of the drift and volatility functions in the Langevin equation, or by introducing a reflecting boundary at $0$ to the associated Fokker-Planck equation. To describe the fluctuating diffusivity by subordination, one needs to obtain the probability density of the integrated diffusivity
\begin{equation}
    Y_t=\int\limits_0^tD_{t'}dt'.
\end{equation} 
The Laplace transform of the probability density of $Y_t$ conditioned on the value $D_0$ of $D_t$ at $t=0$,
\begin{equation}
\hat{Q}(s,t|D_0)=\mathbb{E}\left[\exp\left(-s\int\limits_{0}^tD_{t'}dt'\right) \Bigg|D_0\right],
\end{equation}
satisfies the Feynman-Kac formula which for the considered case of time-independent coefficients $a_D$ and $b_D$ can be written via time reversal as \cite{Grebenkov2018}
\begin{equation}
\left[-\frac{\partial}{\partial t}+a_D(D_0)\frac{\partial}{\partial D_0}+\frac{1}{2}b_D^2(D_0)\frac{\partial^2}{\partial D_0^2}-sD_0\right]\hat{Q}(s,t|D_0)=0,
\end{equation}
subject to the condition $\hat{Q}(s,t=0|D_0)=1$. The Laplace transform of the probability density of the integral of the CIR process $\hat{Q}(D,t|D_0)$ was named $\Upsilon(D,t|D_0)$ in our previous paper \cite{Lanoiselee2018b}; here we keep the notation $\Upsilon(D,t|D_0)$ for the Laplace transform of the probability density of any integrated diffusivity process that will be used in the context of first-passage time statistics in Sec. \ref{sec:first_passage_bounded_domain}. The marginal probability density is obtained by averaging the initial diffusivity $D_0$ over the stationary distribution of diffusivity $\Pi(D_0)$
\begin{equation}\label{eq:prob_density_integ_diff_diff}
\hat{Q}(s,t)=\int\limits_0^\infty \Pi(D_0)\hat{Q}(s,t|D_0)\, dD_0.
\end{equation}

Assuming the Gaussian form (\ref{eq:gaussian_disp_distrib}) of $p(x,u)$ in Eq. (\ref{eq:diff_diff_subor}) and performing the Fourier transform with respect to $x$, one gets
\begin{equation}
\tilde{P}(k,t)=\int_{-\infty}^\infty e^{ikx}P(x,t)dx=\int_0^\infty e^{-\mu_x^2k^2u}Q(u,t)du = \hat{Q}(s=\mu_x^2k^2,t).    
\end{equation}


\subsection{Superposition of independent processes}
Either when the observed dynamics is sampled from a collection of independent particles with a constant but random diffusion coefficient, or when describing the particles motion in a spatially heterogeneous medium within a displacement range smaller than the correlation length of disorder, an appropriate description is obtained through a superposition of processes which in the former case could be called heterogeneous ensemble of non-Gaussian particles \cite{Sliusarenko2019} while the latter case is termed superstatistics \cite{Beck2003,Beck2005,Chechkin2017,Lanoiselee2018a}. The propagator of the ensemble of particles, $P(x,t)$, is a superposition of individual propagators $p(x,t|D)$ with diffusion coefficients picked from a probability density $\Pi(D)$. When the spatial coordinates of the individual propagator $p(x,t|D_\eta)$ scale as $\propto \sqrt{D_\eta t^\eta}$ (i.e. anomalous diffusion) with $\eta\in[0,2]$ being the anomalous exponent and $D_\eta$ the generalized diffusion coefficient, then the propagator can be written $p(x|D_\eta t^\eta)$. The superimposed propagator,
\begin{equation}
    P(x,t)=\int\limits_0^\infty p(x|D_\eta t^\eta)\Pi(D_\eta)dD_\eta,
\end{equation}
can be reformulated in terms of a subordination integral through the variable change $u=D_\eta t^\eta$ 
\begin{equation} \label{eq:superstat_subor}
P(x,t)=\int\limits_0^\infty p(x,u)R(u,t)du,
\end{equation}
 where the probability density of integrated diffusivity is
\begin{equation}\label{eq:superstat_subordinator}
    R(u,t)=\frac{1}{t^\eta}\Pi\left(\frac{u}{t^\eta}\right).
\end{equation}

\section{Continuous-time random integrated diffusivity (CTRID)}
\label{sec:CTR_integrated_diffusivity}

The probability density of integrated diffusivity in the case of diffusing diffusivity presented Sec. \ref{sec:integrated_diffusing_diffusivity_description} can be difficult to handle due to the intricate time dependence of the propagator. In order to be able to combine fluctuating diffusivity with other models of heterogeneity, we propose a simple model of integrated diffusivity that reproduces the main features of diffusing diffusivity. Instead of describing the fluctuations of diffusivity, we directly model its integral. We model the integrated diffusivity by a piecewise constant function which takes increasing random values on successive time intervals of random durations.
We define the {\it transit times} to be the random time spent between two successive increments of integrated diffusivity. We consider both the transit times and the increments of integrated diffusivity to be independent, although this assumption can be relaxed (see \cite{Chechkin2009}). While the piecewise constant integrated diffusivity and a continuously increasing one are different at short times, their long-time behaviors are expected to be the same. Using the renewal theory, we construct a formula for the probability density of integrated diffusivity $R(u,\tau)$ (see Eq. (\ref{eq:general_subordination})). Then, based on the long-time behavior of our version of the Montroll-Weiss formula, we establish an advection equation and a fractional advection equation that describe long-time and large-integrated-diffusivity-increments behavior of the CTRID that we solve and apply in next sections.

\begin{center}
\begin{figure}[h!]
 \includegraphics[width=\textwidth]{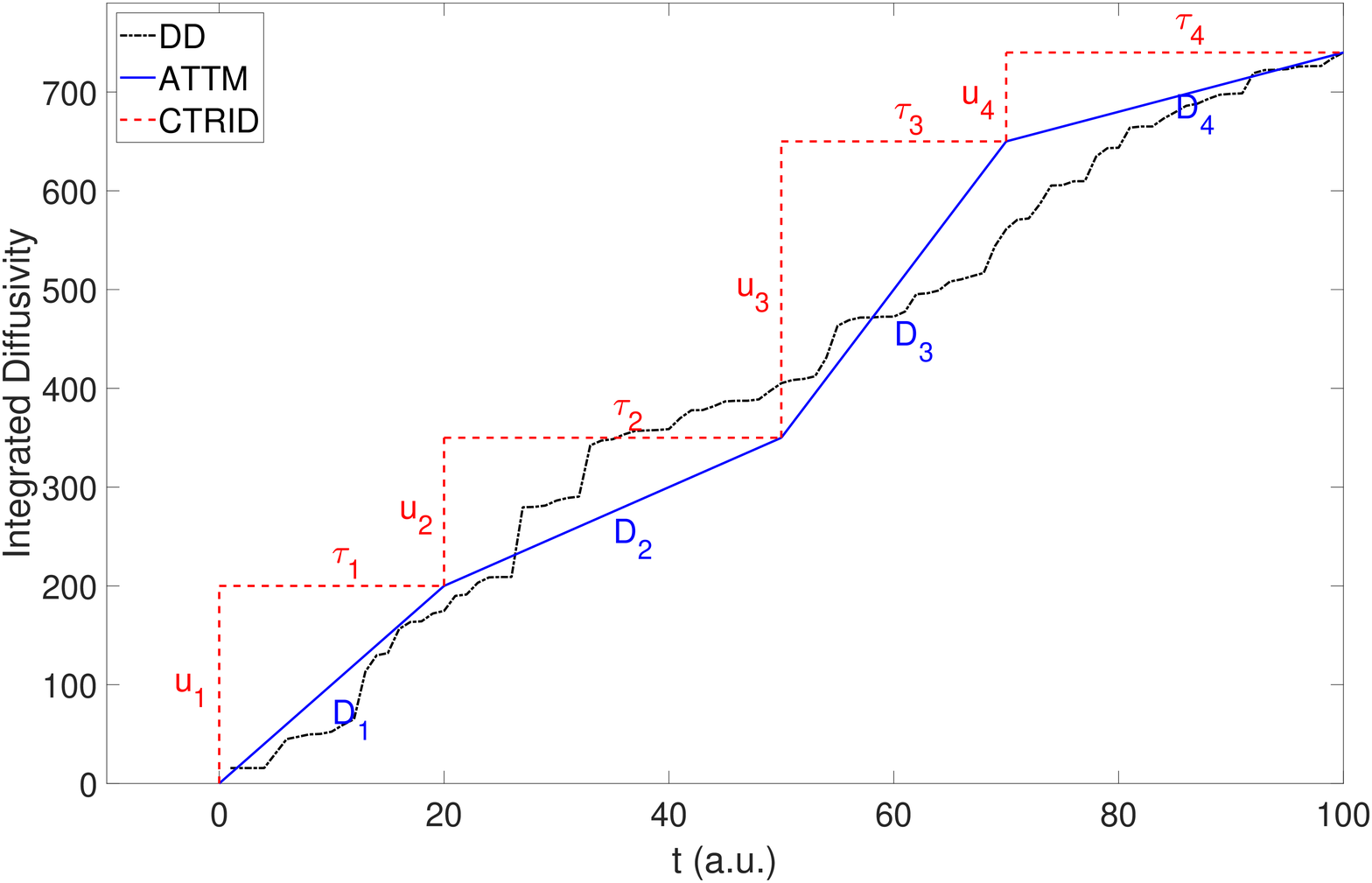}
   \caption{Illustration of the continuous-time random integrated diffusivity (red dashed line) where the integrated diffusivity takes piecewise constant values $u_1,u_2,\ldots,u_n$ during random times $\tau_1,\tau_2,\ldots,\tau_n$. The black dash-dotted line shows the annealed transit time model where the diffusion coefficient $D_1,D_2,\ldots,D_n$ remains constant during random times $\tau_1,\tau_2,\ldots,\tau_n$. Finally the blue line presents the continuously increasing integrated diffusing diffusivity as a function of time (arbitrary units).}
  \label{fig:cmp_ATTM_CTRID}
\end{figure}
\end{center}

We also note that our CTRID model is different from the annealed-transit-time model in which the diffusivity is piecewise constant \cite{Massignan2014}. Figure \ref{fig:cmp_ATTM_CTRID} illustrates a random path of integrated diffusivity in diffusing diffusivity model (DD), ATTM model and our CTRID model, suggesting that all these models should be equivalent in the long-time limit. In the two latter cases, the parameters of the models can be tuned to be much closer to the DD model than presented here.

\subsection{Renewal formula}
We obtain the distribution $R(u,\tau)$ by adapting the continuous-time random walk framework. 
We choose a stationary strictly positive distribution of integrated diffusivity jumps with density $r(u)$ and a stationary strictly positive distribution of transit times with density $\phi(t)$.
The probability density for an integrated diffusivity to be equal to $u$ after $m$ jumps is
\begin{equation}
r_m(u)=\int\limits_{0}^u r_{m-1}(u-u')r(u')du',
\end{equation}
which in the Laplace domain becomes 
\begin{equation}
    \hat{r}_m(s_u)=\left[\hat{r}(s_u)\right]^m.
\end{equation}
In the same way, one obtains the probability density of making $m$ integrated diffusivity increments in a time $\tau$ in the Laplace domain
\begin{equation}
\hat{\phi}_m(s_\tau)=\left[\phi(s_\tau)\right]^m,
\end{equation}
whereas the probability of no integrated diffusivity change for a time $\tau$ reads in the Laplace domain as 
\begin{equation}
  \hat{\Phi}_{0}(s_\tau)=\frac{1-\hat{\phi}(s_\tau)}{s_\tau}.
\end{equation}
Summing over all possible numbers of jumps, one gets for the double Laplace transform of $R(u,\tau)$:
\begin{equation}
 \hat{\hat{R}}(s_u,s_\tau)=\sum_{m=0}^\infty \hat{r}_m(s_u)\hat{\phi}_m(s_\tau)\hat{\Phi}_0(s_\tau),
\end{equation}
which simplifies to an analog of Montroll-Weiss formula
 \begin{equation}\label{eq:montroll_weiss}
 \hat{\hat{R}}(s_u,s_\tau)=\frac{1-\hat{\phi}(s_\tau)}{s_\tau}\frac{1}{1-\hat{r}(s_u)\hat{\phi}(s_\tau)}.
 \end{equation}
For instance, when $\hat{r}(s_u)$ is a Gamma distribution, one obtains the integral of a time-delayed version of the Gamma process studied in \cite{Kumar2018a}.
\subsection{Advection equation}

We first consider the case when both transit times and increments of integrated diffusivity have finite means $\mu_\tau$ and $\mu_u$ respectively. For small $s_\tau$ and $s_u$, one has 
\begin{equation}
\hat{\phi}(s_\tau)=1-\mu_\tau s_\tau+\ldots
\end{equation}
and 
\begin{equation}
\hat{r}(s_u)=1-\mu_u s_u+\ldots,
\end{equation}
 so that
 \begin{equation}
  \hat{\hat{R}}(s_u,s_\tau)\approx\frac{\mu_\tau}{\mu_u s_u+\mu_\tau s_\tau}.
 \end{equation}
Using the properties of the Laplace transform we obtain that in the long-transit-time and large-integrated-diffusivity-jump limit, $R(u,\tau)$ is described by an advection equation
\begin{equation}
\frac{\partial}{\partial \tau}R(u,\tau)=-\frac{\mu_u}{\mu_\tau}\frac{\partial}{\partial u}R(u,\tau).
\end{equation}

Given an initial distribution $R_0(u)$ that verifies $R_0(u)=0$ for $u<0$, the solution is
\begin{equation}\label{eq:sol_advec_eq}
R(u,\tau)=R_0\left(u-\frac{\mu_u}{\mu_\tau}\tau\right).
\end{equation}
The functional form $R_0(z)$ of the distribution is preserved but the mode of the distribution has a constant drift over time. In this way, $u$ is strictly increasing and it is thus an appropriate subordinator. The particular choice $R_0(z)=\delta(z)$ makes the integrated diffusivity deterministic, implying a constant diffusion coefficient. The integrated diffusivity is characterized by both a diffusivity scale $\mu_u$ representing the strength of integrated diffusivity fluctuations, and a timescale $\mu_\tau$. 

We illustrate the behavior of this model by considering a simple case when $R_0(u)$ is an exponential distribution 
\begin{equation}\label{eq:def_Advection_case_R0}
R_0(z)=\frac{1}{\mu_u}e^{-u/\mu_u}\Theta(u),
\end{equation}
with the integrated diffusivity scale $\mu_u$, and $\Theta(u)$ is the Heaviside function.
The Laplace transform of $R(u,t)$ with respect to $u$ is
\begin{equation}
\hat{R}(s_u,t)=\frac{1}{1+\mu_us_u}\exp\left(-\frac{\mu_u}{\mu_\tau}s_u\tau\right).
\end{equation}
Using Eq. (\ref{eq:gaussian_disp_distrib}) for $p(x,u)$ we deduce the characteristic function
\begin{equation}\label{eq:sol_Advection_case_R0}
\tilde{P}(k,t)=\frac{1}{1+\mu_x^2 k^2}\exp\left(-\frac{\mu_x^2}{\mu_t}k^2t\right).
\end{equation}
The first factor is the characteristic function of the Laplace distribution while the second factor is that of the Gaussian distribution. When $\mu_x^2 k^2\ll 1$ the distribution is well approximated by a Gaussian distribution, while when $\frac{\mu_x^2}{\mu_t}t\ll 1$, it is well approximated by a Laplace distribution. When $t/\mu_t\gg 1$, the distribution is Gaussian with exponential tails. However, for sufficiently long time, the exponential tails cannot be observed in practice due to a limited amount of measurable data. Note that this is exactly the propagator for Brownian motion with additional Laplace distributed measurement noise.

\begin{center}
\begin{figure}[h!]
 \includegraphics[width=\textwidth]{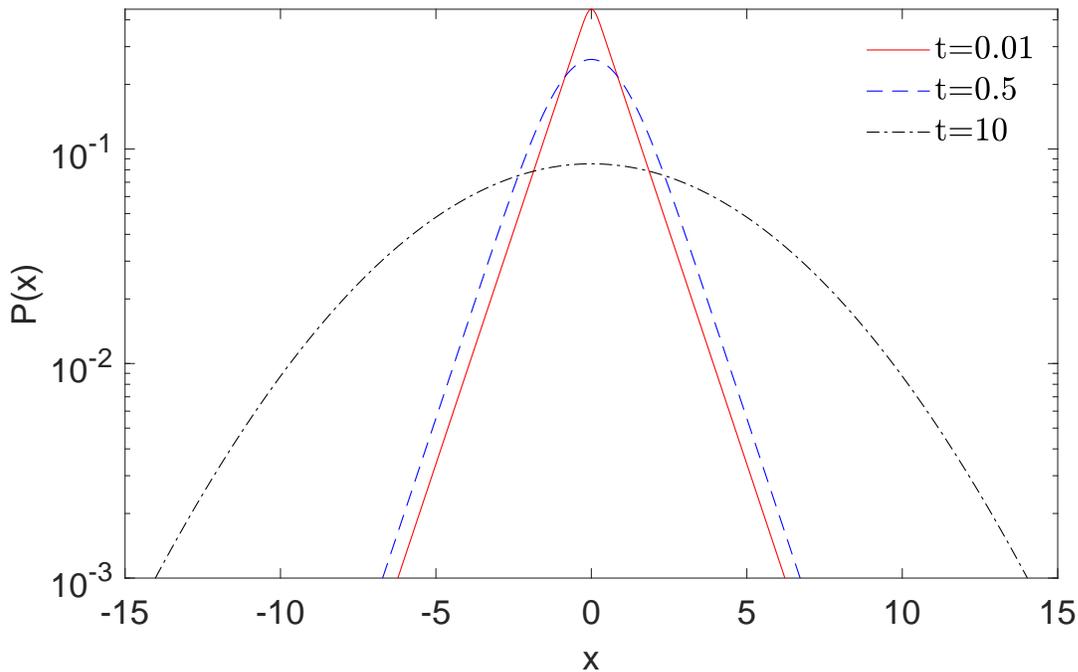}
   \caption{Propagator from Eq. (\ref{eq:CTRID_propag_finite}) for a CTRID model with Gaussian displacements with parameters $\mu_x=1$ and $\mu_t=1$ at time $t=0.01$ (red line), $t=0.5$ (blue dashed line) and $t=10$ (black dashed-dotted line). }
  \label{fig:advection_solution}
\end{figure}
\end{center}

The propagator in real space is then obtained through the convolution integral
\begin{equation}
P(x,t)=
\frac{1}{2\mu_x^2\sqrt{2t/\mu_t}}
\int\limits_{-\infty}^\infty 
\exp\left(-\frac{|x'|}{\mu_x}\right)
\exp\left(-\frac{(x-x')^2}{4\mu_x^2t/\mu_t}\right)
dx',
\end{equation}
which yields
\begin{eqnarray}\label{eq:CTRID_propag_finite}\nonumber
P(x,t)&=&
\frac{1}{2\mu_x}\sqrt{\frac{\pi}{2}}
\exp\left(\frac{t}{\mu_t}-\frac{x}{\mu_x}\right)
\left[
\erfc\left(\sqrt{\frac{t}{\mu_t}}\left(1-\frac{x\mu_t}{2\mu_x t}\right)
\right)
\right.\\&&\qquad+\left.
\exp\left(\frac{2x}{\mu_x}\right)\erfc\left(\sqrt{\frac{t}{\mu_t}}\left(1+\frac{x\mu_t}{2\mu_x t}\right)\right)\right],
\end{eqnarray}
where $\erfc(x)$ is the complementary error function. Figure \ref{fig:advection_solution} illustrates the propagator from Eq. (\ref{eq:CTRID_propag_finite}) for CTRID with Gaussian distributed displacements. As expected, the propagator is close to a Laplace probability density function at short time and then converges to a Gaussian one a longer time.
Similar calculations can be conducted for more general $R_0(u)$ (e.g. generalized Gamma distribution \cite{Sposini2018}).

Because the distribution of the integrated diffusivity preserves its shape at all times, the mean squared displacement, as well as all even moments of displacements, are not zero at time $t=0$. 
For this reason, we characterize the distribution by its moments minus their initial values.
In this case the mean squared displacement is a linear function of time,
\begin{equation}
\langle X^2(t)\rangle-\langle X^2(0) \rangle=2\frac{\mu_x^2}{\mu_t}t,
\end{equation}
and the fourth moment is a quadratic function of time
\begin{equation}
\langle X^4(t) \rangle-\langle X^4(0) \rangle=12\left(\frac{\mu_x^2}{\mu_t}t\right)^2+24\frac{\mu_x^4}{\mu_t}t,
\end{equation}
such that the shifted normalized excess kurtosis is
\begin{equation}
\gamma(t)=\frac{1}{3}\frac{\langle X^4(t) \rangle-\langle X^4(0) \rangle}{\left(\langle X^2(t)\rangle-\langle X^2(0)\rangle\right)^2}=2\frac{\mu_t }{t}.
\end{equation}
The long-time behavior is the same as in the diffusing diffusivity model at times longer that the diffusivity autocorrelation time \cite{Lanoiselee2018a}. As this minimalistic model reproduces well the convergence to Gaussian probability density of displacements of a process with uncorrelated fluctuations of diffusivity, it can be considered as a simplified model of 'diffusing diffusivity'.

\subsection{The fractional advection equation}
\label{sec:fractional_advection}
Now we get to the case of heavy-tailed distribution of both transit times and integrated diffusivity jumps. The developments for small $s_\tau$ and small $s_u$ read:
\begin{equation}
  \hat{\phi}(s_\tau)=1-(\mu_\tau s_\tau)^\delta+\ldots
\end{equation}
and
\begin{equation}
    \hat{r}(s_u)=1-(\mu_u s_u)^\gamma+\ldots,
\end{equation}
where $\mu_\tau$ and $\mu_u$ are respective scales, while $0<\delta<1$ and $0<\gamma<1$ are the scaling exponents.
Eq. (\ref{eq:montroll_weiss}) yields thus
\begin{equation}\label{eq:R_su_stau}
\hat{\hat{R}}(s_u,s_\tau)\approx\frac{\mu_\tau^\delta s_\tau^{\delta-1}}{\mu_\tau^\delta s_\tau^{\delta}+\mu_u^\gamma s_u^{\gamma}},
\end{equation}
so that $R(u,\tau)$ satisfies the fractional advection equation
\begin{equation}\label{eq:frac_advec_eq}
 \frac{\partial}{\partial\tau}R(u,\tau)=-\frac{\mu_u^\gamma}{\mu_\tau^\delta}~_0\mathcal{D}_\tau^{1-\delta}\left[\mathcal~_0{D}_u^\gamma R(u,\tau)\right],
\end{equation}
where $~_0\mathcal{D}_\tau^{1-\delta}$ and $~_0\mathcal{D}_\tau^{\gamma}$ are the Riemann-Liouville operators, e.g. 
\begin{equation}
    ~_0\mathcal{D}_\tau^{1-\delta}f(\tau)= \frac{1}{\Gamma(\delta)}\frac{\partial}{\partial\tau} \int\limits_{0}^\tau \frac{1}{(\tau-\tau')^{1-\delta}}f(\tau') d\tau'.
\end{equation}
The inverse Laplace transform of Eq. (\ref{eq:R_su_stau}) with respect to $s_\tau$ yields \cite{Haubold2011}
\begin{eqnarray}\label{eq:R_su_tau}
\tilde{R}(s_u,\tau) &=& E_\delta \left(-\frac{\mu_u^\gamma}{\mu_\tau^\delta}  s_u^{\gamma}\tau^{\delta}\right),
\end{eqnarray}
where $E_\delta(z)$ is the Mittag-Leffler function.
The inverse Laplace transform of Eq. (\ref{eq:R_su_tau}) with respect to $s_u$ can be expressed in terms of the Fox H function \cite{Mathai2010}
\begin{equation}\label{eq:CTRD_solution}
R(u,\tau)=\frac{1}{u}H_{2,2}^{1,1}\left[\frac{u^{\gamma}\mu_\tau^\delta}{\mu_u^\gamma\tau^{\delta}}  \middle\vert
\begin{array}{c}
(1,1),(1,\delta)\\
(1,1),(1,\gamma) 
\end{array}
\right].
\end{equation}

\section{Multiple subordination}
\label{sec:multiple_subordination}
In this section we discuss the possibility of describing time-dependent heterogeneities of mixed origins. The Langevin equation associated with time-dependent diffusivity and waiting times between displacements can be written 
as a system of three independent Langevin equations
\begin{equation}\label{Langevin_Equation}
\left\{
    \begin{array}{ll}
\displaystyle \frac{dx}{d u}= \sqrt{2}\,d\eta(u) \\
\displaystyle \frac{d u}{d\tau}=D(\tau),\\
\displaystyle \frac{d\tau}{dt}=S(t),
    \end{array}
\right.
\end{equation}
where $d\eta(u)$ is a continuous noise term (e.g. increments of a L\'{e}vy process) with a correlation function $\langle d\eta(u)\,d\eta(u')\rangle=\delta(u-u')$, $D(\tau)$ is the time-dependent diffusivity, and $S(t)$ describes the increments of the internal time; both $D(\tau)$ and $S(t)$ can be stochastic processes. There are at least two methods to solve these equations. The first method is to consider the three-dimensional Fokker-Planck equation for the joint probability density $P(x,u,\tau,t)$. The marginal propagator $P(x,t)$ that can be estimated from an experiment, is then obtained by averaging over $u$ and $\tau$
 \begin{equation}
     P(x,t)=\int\limits_0^\infty\int\limits_0^\infty P(x,u,\tau,t)\,du\,d\tau.
 \end{equation} 
We choose the second method consisting in solving three independent Fokker-Planck equations: (i) one on the probability density $p(x,u)$ of displacements, (ii) the other on the probability density $R(u,\tau)$ of the integrated diffusivity and (iii) the last one describing the evolution of the probability density $T(\tau,t)$ of the internal time of waiting times. The mathematical key is that a subordinated subordinator is itself a subordinator. Therefore, the probability density obtained by successively subordinating Brownian motion with different subordinators can be seen  as Brownian motion subordinated by a single more sophisticated subordinator. For example, the pathwise properties of a particle with Gaussian displacements with diffusing diffusivity and waiting times between jumps can be represented as
\begin{equation}
    x(t)=y[I(S(t))],
\end{equation}
with the standard Brownian motion $y(t)$ and the subordinator $I(t)$ interpreted as integrated diffusivity
\begin{equation}
    I(t)=\inf\left\lbrace \tau>0:\frac{\mu_t}{\mu_x^2}\int\limits_0^\tau D_sds>t\right\rbrace.
\end{equation}
In turn $S(t)$ is the subordinator associated to waiting times, defined in Eq. (\ref{eq:def_subor_CTRW}), whose distribution is given in Eq. (\ref{eq:CTRW_subordinator}). Let us define $x(t)=y\star S_1\star S_2\star (t)$ as the operation associated to a double subordination $x(t)=y[S_1(S_2(t))]$.
 In general, the subordination can be done an arbitrary number of times, and the characteristic function of the $n$-fold subordinated process
\begin{equation}
x(t)=y\star S_1\star S_2\star\ldots S_n\star (t),
\end{equation}
with $y(t)$ distributed according to $p(x,u)$, is
\begin{equation}\label{eq:general_subor_char_fun} 
\fl
P(x,t)=\int\limits_0^\infty\int\limits_0^\infty\ldots \int\limits_0^\infty p(x,u_1)R_1(u_1,u_2)R_2(u_2,u_3)\ldots R_n(u_n,t)\,du_1du_2\ldots du_n    
\end{equation}
with the probability density $R_i$ of the $i$-th subordinator.

In this section we study the propagator for processes with combined time-fluctuating diffusivity and waiting times between jumps by combining Eq. (\ref{eq:CTRW_subor}) and Eq. (\ref{eq:diff_diff_subor}) into a double subordination formula 
\begin{equation}\label{eq:general_subordination}
P(x,t)= \int\limits_0^\infty \int\limits_0^\infty  p(x,u)  R_1(u,\tau)R_2(\tau,t)\,du\, d\tau.
\end{equation}
This representation allows one to model a great variety of diffusion phenomena by adjusting $p(x,u)$, $R_1(u,\tau)$ and $R_2(\tau,t)$.

In Table \ref{table:List_Equations}, we list some examples of diffusion processes that can be described within the double subordination framework. In each case we present the characteristic function of displacement $\tilde{p}(k,u)$, and the probability densities $R_1(u,\tau)$ and $R_2(\tau,t)$ of two subordinators, necessary to get the characteristic function of the desired mixed non-Gaussian model. Note that the double subordination can also be used to model phenomena with two-power-law relaxation responses such as dielectric breakdown \cite{Weron2010,Weron2017}.

\begin{sidewaystable}
\begin{tabular}{|c|c|c|c|}
\hline
  Model & $\displaystyle \tilde{p}(k,u)$ & $\displaystyle R_1(u,\tau)$ & $\displaystyle R_2(\tau,t)$ \\
   \hline
   & & & \\
  Brownian motion & $\displaystyle \exp\left(-k^2\mu_x^2 u/\mu_u\right)$ & $\displaystyle \delta(u-\mu_u \tau/\mu_\tau)$ & $\displaystyle \delta\left(\tau-\mu_\tau t/\mu_t\right)$  \\   & & & \\
 
 Random diffusivity (CTRID)& $\displaystyle \exp\left(-k^2\mu_x^2 u/\mu_u\right)$ & $\displaystyle R_0\left(u-\frac{\mu_u\tau}{\mu_\tau}\right)\Theta\left(u-\frac{\mu_u\tau}{\mu_\tau}\right)$ & $\displaystyle \delta\left(\tau-\mu_\tau t/\mu_t\right)$ \\   & & & \\
 
 \makecell{Superposition of \\ Brownian motions} & $\displaystyle \exp\left(-k^2\mu_x^2u/\mu_u\right)$ &  $\displaystyle \frac{1}{\tau}\Pi\left(\frac{u}{\tau}\right)$ & $\displaystyle \delta\left(\tau-\mu_\tau t/\mu_t\right)$ \\   & & & \\
 
  Diffusing diffusivity & $\displaystyle \exp\left(-k^2\mu_x^2u/\mu_u\right)$ & $\displaystyle Q(u,\tau)$ & $\displaystyle \delta\left(\tau-\mu_\tau t/\mu_t\right)$ \\   & & & \\
\makecell{ Superposition of \\
 diffusing diffusivities} & $\displaystyle \exp\left(-k^2\mu_x^2u/\mu_u\right)$ & $\displaystyle Q\left(u,\tau\right)$ &  $\displaystyle \frac{1}{t}\Pi\left(\frac{\tau}{t}\right)$ \\   & & & \\
 
 \makecell{ Space-fractional \\
 diffusion (Symmetric) }
 & $\displaystyle \exp\left(-|k|^\alpha\mu_x^\alpha u/\mu_u\right)$ & $\displaystyle \delta(u-\mu_u \tau/\mu_\tau)$ & $\displaystyle \delta\left(\tau-\mu_\tau t/\mu_t\right)$  \\   & & & \\
 
  Time-fractional diffusion & $\displaystyle \exp\left(- k^2\mu_x^2u/\mu_u\right)$ & $\displaystyle \delta(u-\mu_u \tau/\mu_\tau)$ & $\displaystyle \frac{1}{\beta \tau }H_{1,1}^{1,0}\left[\frac{\mu_t^\beta\tau}{t^\beta\mu_\tau}\middle\vert
\begin{array}{cc}
(0,\beta)\\(0,1) 
\end{array}
\right]$  \\   & & & \\
 
 Diffusivity-fractional diffusion & $\displaystyle \exp\left(-k^2\mu_x^2u/\mu_u\right)$ & $\displaystyle \frac{1}{u}H_{2,2}^{1,1}\left[\frac{u^{\gamma}\mu_\tau^\delta}{\mu_u^\gamma\tau^{\delta}} \middle\vert
\begin{array}{c}
(1,1),(1,\delta)\\
(1,1),(1,\gamma) 
\end{array}
\right]$& $\displaystyle \delta\left(\tau-\mu_\tau t/\mu_t\right)$  \\   & & & \\

\makecell{   Space-diffusivity-time\\ fractional diffusion} & $\displaystyle \exp\left(-|k|^\alpha\mu_x^\alpha u/\mu_u\right)$ 
   & 
   $\displaystyle \frac{1}{u}H_{2,2}^{1,1}\left[\frac{u^{\gamma}\mu_\tau^\delta}{\mu_u^\gamma\tau^{\delta}} \middle\vert
\begin{array}{c}
(1,1),(1,\delta)\\
(1,1),(1,\gamma) 
\end{array}
\right]$  
& 
$\displaystyle \frac{1}{\beta \tau }H_{1,1}^{1,0}\left[\frac{\mu_t^\beta\tau}{t^\beta\mu_\tau}\middle\vert
\begin{array}{cc}
(0,\beta)\\(0,1) 
\end{array}
\right]$ \\   & & & \\
\hline
\end{tabular}
\caption{Summary of models of non-Gaussian diffusion of mixed origins that can be obtained by double subordination. $R_0(u)$ is the distribution of integrated diffusivity obtained in Eq. (\ref{eq:sol_advec_eq}), $\Pi(z)$ is the distribution of diffusion coefficients distributed across a population of diffusing particles (see Eq.(\ref{eq:superstat_subordinator})), $Q\left(u,\tau\right)$ is the probability density of integrated diffusing diffusivity (see Eq.(\ref{eq:prob_density_integ_diff_diff})), the seventh and the last rows use for $R_2(\tau,t)$ the probability density of the inverse stable subordinator defined in Eq (\ref{eq:CTRW_subordinator}) and the two last rows use for $R_1(u,\tau)$ the solution to the fractional advection equation (\ref{eq:CTRD_solution}).}
\label{table:List_Equations}
\end{sidewaystable}


\subsection{Superposition of processes with Continuous Time Random Integrated Diffusivity}

When analyzing single particle data from an experiment on living cells, a representative statistical set is often obtained from numerous cells. If we assume that the dynamics of an individual tracer can be described by a diffusing-diffusivity model, then each tracer may have a different long-term average diffusion coefficient due to the cell-to-cell variability. Then the propagator obtained from the ensemble of acquired trajectories is a superposition of diffusing diffusivity processes. Taking the CTRID model as a example of diffusing diffusivity process, our approach allows for the first time to describe such superposition:
\begin{equation}
    \tilde{P}(k,t)=\int\limits_0^\infty\int\limits_0^\infty \tilde{p}(k,u)\,R_0\left(u-\frac{\mu_u\tau}{\mu_\tau}\right)\Theta\left(u-\frac{\mu_u\tau}{\mu_\tau}\right)\,\frac{1}{t}\Pi\left(\frac{\tau}{t}\right)\,du\,d\tau.
\end{equation}
Setting $\tilde{p}(k,u)=\exp\left(-k^2\mu_x^2u/\mu_u\right)$, $R_0(u)$ to be defined in Eq. (\ref{eq:def_Advection_case_R0}), and the distributed diffusion coefficient across the population of cells to follow a Gamma distribution 
\begin{equation}
\Pi\left(z\right)=\frac{z^{\nu-1}\mu_t^\nu}{\Gamma(\nu)\mu_\tau^\nu} \exp\left(-\frac{z\mu_t}{\mu_\tau}\right),
\end{equation}
with dimensionless scale $\mu_t/\mu_\tau$ and shape parameter $\nu$, the characteristic function is
\begin{equation}
\tilde{P}(k,t)=\frac{ 1}{\left(1+\mu_x^2k^2\right)\left(1+\mu_x^2k^2t/\mu_t\right)^{\nu}}.
\end{equation}
At this stage, we notice that the initial non-Gaussian propagator converges to another non-Gaussian propagator in the long-time limit. The cross over between two non-Gaussian regimes is an important feature that only arises when at least two heterogeneity sources are present.

The distribution is then the convolution of a Laplace distribution with the probability density (16) from \cite{Lanoiselee2018a}
\begin{equation}
\fl
    \mathcal{F}^{-1}\left[\frac{ 1}{\left(1+\frac{\mu_x^2}{\mu_t}k^2t\right)^{\nu}}\right]=
    \frac{2^{1/2-\nu}}{\Gamma(\nu)}\sqrt{\frac{\nu\mu_t}{\pi\mu_x^2t}}
    \left(\frac{|x|}{\mu_x}\sqrt{\frac{\nu\mu_t}{t}}\right)^{\nu-1/2}K_{\nu-1/2}\left(\frac{|x|}{\mu_x}\sqrt{\frac{\nu\mu_t}{t}}\right),
\end{equation} 
where $K_{\nu}(z)$ is the modified Bessel function of the second kind.

\begin{center}
\begin{figure}[h!]
 \includegraphics[width=\textwidth]{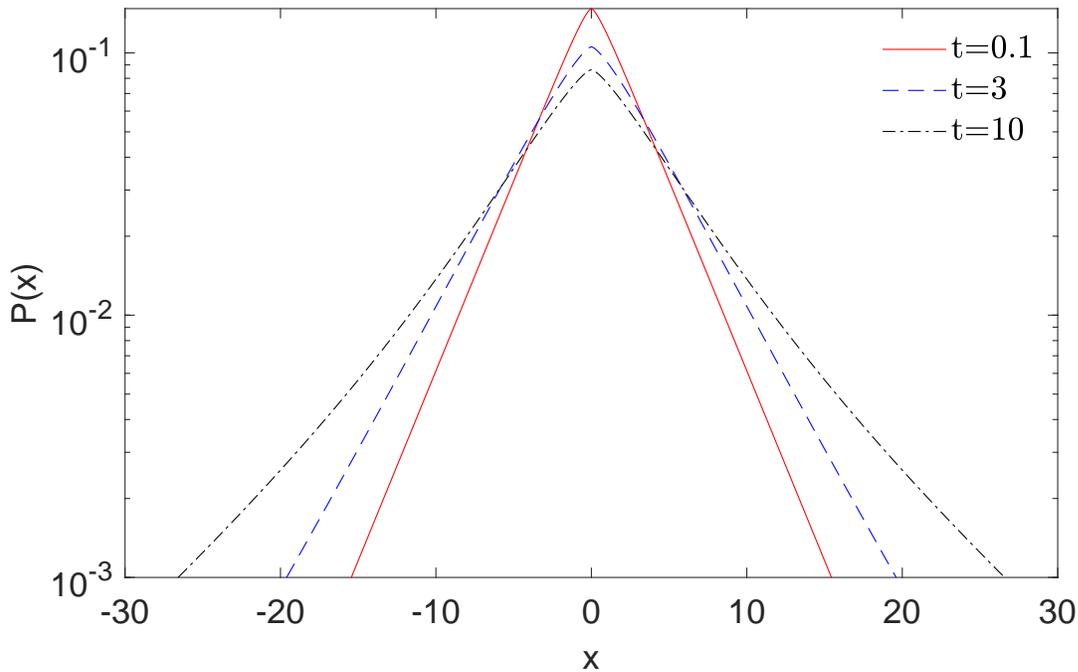}
   \caption{The propagator of a superposition of processes with Gaussian displacements and integrated diffusivity modeled by CTRID obtained from Eq. (\ref{eq:propagator_supertat_CTRID}) with parameters $\mu_x=3$, $\mu_t=1$ and $\nu=0.3$ at time $t=0.1$ (red line), $t=3$ (blue dashed line) and $t=10$ (black dashed-dotted line). }
  \label{fig:superstat_CTRID_solution}
\end{figure}
\end{center}

The propagator in real space can be computed through the convolution integral 
\begin{eqnarray}\label{eq:propagator_supertat_CTRID}\nonumber
    P(x,t)&=&
    \frac{2^{1/2-\nu}}{\mu_x^2\Gamma(\nu)}\sqrt{\frac{\nu\mu_t}{\pi t}}
    \int_{-\infty}^\infty
    \exp\left(-\frac{|x-x'|}{\mu_x}\right)
    \left(\frac{|x'|}{\mu_x}\sqrt{\frac{\nu\mu_t}{t}}\right)^{\nu-1/2}\\
&&\qquad\qquad\qquad\qquad\times    K_{\nu-1/2}\left(\frac{|x'|}{\mu_x}\sqrt{\frac{\nu\mu_t}{t}}\right)dx',
\end{eqnarray}
which in the case $\nu=1$ reduces to 
\begin{equation}
    P(x,t)=
    \frac{1}{\mu_x\left(\frac{t}{\mu_t}-1\right)}
    \sqrt{\frac{2}{\pi}}
    \left[
    \sqrt{\frac{t}{\mu_t}}\exp\left(-\frac{|x|}{\mu_x}\sqrt{\frac{\mu_t}{t}}\right)-\exp\left(-\frac{|x|}{\mu_x}\right)\right],
     \end{equation}
which remains positive and properly normalized at all times.
Figure \ref{fig:superstat_CTRID_solution} illustrates the propagator from Eq. (\ref{eq:propagator_supertat_CTRID}), which behaves at short times as a Laplace distribution and then converges to an even less Gaussian distribution at long time (the kurtosis has increased).
\subsection{Space-diffusivity-time fractional diffusion}
\label{sec:SDT_fractional_diffusion}
Diffusion processes with a stable distribution of particle jumps (space fractional diffusion) and power law waiting times between jumps (time fractional diffusion) have been investigated \cite{Metzler2000,Mainardi2001,Meerschaert2002,Chen2012,Pagnini2016}.
Here our approach allows us to generalize these processes by introducing an example of non-Gaussian diffusion in which four mechanisms leading to anomalous diffusion are superimposed that we call ``space-diffusivity-time fractional diffusion''.

The jumps are distributed according to a symmetric stable distribution and exhibit fluctuating diffusivity for which the integral is modeled according to the fractional advection equation developed in Sec. \ref{sec:fractional_advection}. Finally there are power law waiting times between displacements of the particle similar to that of the time-fractional diffusion model. The system of integro-differential equations describing the whole system is

\begin{subequations}
\begin{align}
&
\label{eq:partial_diff_fractional_displ}
\begin{aligned}
\displaystyle \frac{\partial}{\partial u}p(x,u)=\frac{\mu_x^\alpha}{\mu_u}\frac{\partial^\alpha}{\partial |x|^\alpha}p(x,u), 
\end{aligned}
\\
&
\label{eq:partial_diff_fractional_ID}
\begin{aligned}
\displaystyle \frac{\partial}{\partial\tau}R(u,\tau)=-\frac{\mu_u^\gamma}{\mu_\tau^\delta} ~_0\mathcal{D}_\tau^{1-\delta}\left[~_0\mathcal{D}_u^\gamma R(u,\tau)\right]
\end{aligned}
\\
&
\label{eq:partial_diff_fractional_time}
\begin{aligned}
\displaystyle ~_0\mathcal{D}_t^{\beta} T(\tau,t)=-\frac{\mu_\tau}{\mu_t^\beta}\frac{\partial}{\partial \tau}  T(\tau,t),
\end{aligned}
\end{align}
\end{subequations}
\noindent where the first equation describes diffusion with symmetric stable displacements, the second one is the fractional advection equation for the integrated diffusivity, and the last one is the time-fractional advection equation of the inverse stable subordinator \cite{Meerschaert2013}. 
The characteristic function (i.e., the  Fourier transform with respect to space coordinate of the solution to Eq. (\ref{eq:partial_diff_fractional_displ})),
\begin{equation}\label{eq:displacement_levy_symmetric}
\tilde{p}(k,u)=\exp\left(-\frac{\mu_x^\alpha}{\mu_u}|k|^\alpha u\right)
\end{equation}
is first combined with the probability density of integrated diffusivity from Eq. (\ref{eq:CTRD_solution}) such that
\begin{equation}
\tilde{V}(k,\tau)=\int\limits_0^\infty \tilde{p}(k,u)R(u,\tau) \,du
\end{equation}
is the Laplace transform of $R(u,\tau)$ with respect to $u$ with parameter $s_u=\mu_x^\alpha|k|^\alpha/\mu_u$:
\begin{eqnarray}
\tilde{V}(k,\tau)&=&E_\delta\left(-\frac{\mu_x^{\alpha\gamma}}{\mu_\tau^\delta}|k|^{\alpha\gamma}\tau^{\delta}\right),
\end{eqnarray}
that can be written in terms of the Fox H-function
\begin{eqnarray}
\tilde{V}(k,\tau)=H_{1,2}^{1,1}\left[\frac{\mu_x^{\alpha\gamma}}{\mu_\tau^\delta}|k|^{\alpha\gamma}\tau^{\delta}\middle\vert
\begin{array}{c}
(0,1) \\
(0,1),(0,\delta)
\end{array}
\right].
\end{eqnarray}

The probability density $T(\tau,t)$ of the inverse stable subordinator is given by Eq. (\ref{eq:CTRW_subordinator}).
Then the characteristic function of displacements at time $t$ is expressed as  the subordination integral

\begin{eqnarray}
\fl
\tilde{P}(k,t)&=&\frac{1}{\beta }
\int\limits_0^\infty \tau^{-1} H_{1,2}^{1,1}\left[\frac{\mu_x^{\alpha\gamma}}{\mu_\tau^\delta}|k|^{\alpha\gamma}\tau^{\delta}\middle\vert
\begin{array}{c}
(0,1) \\
(0,1),(0,\delta)
\end{array}
\right] H_{1,1}^{1,0}\left[\frac{\mu_t^\beta\tau}{t^\beta\mu_\tau}\middle\vert
\begin{array}{cc}
(0,\beta)\\(0,1) 
\end{array}
\right]d\tau,
\end{eqnarray}
which can be interpreted as a Mellin transform $\int\limits_0^\infty  \tau^{s-1}f(\tau)d\tau$ with parameter $s=0$ of the product of two Fox H functions \cite{Mathai2010}, which is 
\begin{equation}\label{eq:Space_diff_time_frac_char_fun}
\tilde{P}(k,t)=\frac{1}{\beta}H_{2,3}^{1,2}\left[\frac{\mu_x^{\alpha\gamma}}{\mu_t^{\delta\beta}}|k|^{\alpha\gamma}t^{\delta\beta}\middle\vert
\begin{array}{c}
(1,\delta),(0,1) \\
(0,1),(1,\delta\beta),(0,\delta)
\end{array}
\right].
\end{equation}
Then, the inverse Fourier transform is performed by using the properties of the Fox H-function which gives access to the exact expression for the probability density
\begin{equation}\label{eq:exact_propag_space_diff_time_fractional}
\fl
    P(x,t)=\frac{1}{\beta|x|\sqrt{\pi}}H_{3,4}^{3,1}\left[\frac{\mu_t^{\delta\beta}}{t^{\delta\beta}}\left(\frac{|x|}{2\mu_x}\right)^{\alpha\gamma}\middle\vert
\begin{array}{c}
(1,1),(0,\delta\beta),(1,\delta)\\
(1/2,\alpha\gamma/2),(0,\delta),(1,1),(1,\alpha\gamma/2) 
\end{array}
\right].
\end{equation}

 Because $\alpha$ and $\gamma$ appear as a product in the propagator, the changes of fractional integrated diffusivity plays the same role as the stable displacements of a L\'{e}vy flight. In turn, although $\beta$ and $\delta$ appear as a product that scales time, they differ in the arguments of the Fox H function, suggesting new effects from the combination of both mechanisms. The combined effect of  $\beta<1$ and $\delta<1$ is to slow down the dynamics while the combined effect of $\alpha<2$ and $\gamma<1$ is to make more probable large displacements. For the first time, we are able to show that the combination of waiting times between jumps, integrated diffusivity jumps and transit-times between these, when they are all power-law distributed, leads to a scaling distribution that does not change its shape with respect to time.
\section{First-passage times}
\label{sec:first_passage}
We now look into the problem of the first-passage time statistics which is central for the description of reaction kinetics.

\subsection{First-passage time in any bounded domain in $\mathbb{R}^d$}
\label{sec:first_passage_bounded_domain}
In the case of first-passage time in a bounded domain $\Omega$ in $\mathbb{R}^d$, many results are known for Brownian motion. The approach developed in this article was to design an elaborate annealed disorder description of non-Gaussian diffusion which can be shown to be equivalent to a subordinated Brownian motion. As a consequence, as explicited in \cite{Lanoiselee2018b}, the first-passage time statistics can be obtained by applying appropriate subordination integrals to the similar statistics for Brownian motion.
The propagator of the process inside a bounded domain starting at a position $x_0\in\Omega$ is given by
\begin{equation}
P_\Omega(x,t|x_0)=\sum_{m=0}^\infty u_m(x)u_m(x_0) \Upsilon(\lambda_m,t),
\end{equation}
where $\lambda_m$ is the $m$-th eigenvalue and $u_m(x)$ is the corresponding $L_2$-normalized eigenfunction of the Laplace operator $\Delta$, i.e. solutions of the eigenvalue problem for normal diffusion $\Delta u_m+\lambda_m u_m=0$ in $\Omega$ subject to appropriate boundary conditions (e.g. Dirichlet or Neumann). Here, the subscript `$\Omega$' emphasizes that the propagator is defined in this domain.
The function $\Upsilon(\lambda_m,t)$ is the Laplace transform of the probability density of integrated diffuvity in Eq. (\ref{eq:general_subor_char_fun}) given by the relation
\begin{equation}
\fl
\Upsilon(\lambda,t)=\int\limits_0^\infty\int\limits_0^\infty\ldots \int\limits_0^\infty e^{-\lambda u_1}R_1(u_1,u_2)R_2(u_2,u_3)\ldots R_n(u_n,t)\,du_1du_2\ldots du_n,
\end{equation}
which can be written as combined subordination integral with respect to the Laplace transform of the first subordinator
\begin{equation}
\Upsilon(\lambda,t)=\int\limits_0^\infty\ldots \int\limits_0^\infty\hat{R}_1(s_{u_1}=\lambda,u_2)R_2(u_2,u_3)\ldots R_n(u_n,t)\,du_2\ldots du_n.
\end{equation}
Then the probability density of the first-passage time is 
\begin{equation}
\rho(t|x_0)=-\sum_{m=0}^\infty u_m(x_0) \left(\int\limits_{\Omega} u_m(x)dx\right)\,\frac{\partial}{\partial t}\Upsilon(\lambda_m,t).
\end{equation}
The macroscopic flux is the average of this density with respect to the initial concentration of particles $c(x_0)$ in the medium
\begin{eqnarray}
J(t)&=&-\sum_{m=0}^\infty\, \left(\int\limits_{\Omega} c(x_0)u_m(x_0)dx_0\right) \left(\int\limits_{\Omega}u_m(x)dx\right)\,\frac{\partial}{\partial t}\Upsilon(\lambda_m,t).
\end{eqnarray}
This approach gives access to various first-passage time statistics for any combination of time-dependent heterogeneities.

For instance, in the case of Diffusivity-Time Fractional diffusion model which is the solution of Eq. (\ref{eq:Space_diff_time_frac_char_fun}) with $\alpha=2$ to ensure the process to be a subordinated Brownian motion, one gets 
\begin{equation}
\Upsilon(\lambda_m,t)=
\frac{1}{\beta}H_{2,3}^{1,2}\left[\frac{\mu_x^{2\gamma}}{\mu_t^{\delta\beta}}\lambda_m^{\gamma}t^{\delta\beta}\middle\vert
\begin{array}{c}
(1,\delta),(0,1) \\
(0,1),(1,\delta\beta),(0,\delta)
\end{array}
\right].
\end{equation} 
It is important to notice that the exponent $\gamma$ does not affect the spectral properties of the Laplace operator. 
\subsection{First-passage times on a positive half-axis}
The above analysis is not applicable to first-passage times in unbounded domains for which the Laplace operator has a continuous spectrum. In some cases, one can employ symmetries to get explicit results (see \cite{Lanoiselee2018b} for details). Here, we illustrate this possibility for diffusion on a half-line. From the characteristic function of displacements, the survival probability of a particle in $[0,\infty]$ with absorbing boundary located at $x=0$ is
\begin{equation}
S(t|x_0)=\frac{2}{\pi}\int\limits_0^\infty\frac{1}{k}\sin(kx_0)\tilde{P}(k,t)dk.
\end{equation}
When $\tilde{P}(k,t)$ is obtained as a process with successive subordinations with stable symmetric displacement (see Eq. (\ref{eq:displacement_levy_symmetric})), the survival probability is
\begin{equation}
\fl
S(t|x_0)= \frac{2}{\pi}\int\limits_0^\infty\int\limits_0^\infty\left[\int\limits_0^\infty\frac{1}{k}\sin(kx_0)\exp\left(-\frac{\mu_x^\alpha}{\mu_u} k^\alpha u\right)dk\right]\,R(u,\tau)T(\tau,t)\,du\,d\tau,
\end{equation}
which after the integration over $k$ becomes 
\begin{equation} 
\fl
S(t|x_0)= \frac{1}{\sqrt{\pi}}\int\limits_0^\infty\int\limits_0^\infty H_{2,1}^{1,1}\left[\frac{\mu_x^\alpha}{\mu_u}\left(\frac{2}{x_0}\right)^\alpha u
\middle\vert \begin{array}{c}
(1/2,\alpha/2),(1,\alpha/2) \\
(0,1)
\end{array}
\right]R(u,\tau)T(\tau,t) du\,d\tau.
\end{equation}
When $R(u,\tau)$ is the solution (\ref{eq:CTRD_solution}) of the fractional advection equation obtained in Sec. \ref{sec:fractional_advection} and $T(\tau,t)$ is the inverse stable subordinator from Eq. (\ref{eq:CTRW_subordinator}), the calculation of the survival probability consists in two consecutive Mellin transforms of the product of two Fox H functions 
\begin{eqnarray}\nonumber
\fl
S(t|x_0)=\frac{1}{\sqrt{\pi}}
\int\limits_0^\infty\int\limits_0^\infty \frac{1}{u}
H_{2,1}^{1,1}\left[\frac{\mu_x^\alpha}{\mu_u}\left(\frac{2}{x_0}\right)^\alpha u
\middle\vert \begin{array}{c}
(1/2,\alpha/2),(1,\alpha/2) \\
(0,1)
\end{array}
\right]H_{2,2}^{1,1}\left[\frac{u^{\gamma}\mu_\tau^\delta}{\mu_u^\gamma\tau^{\delta}} \middle\vert
\begin{array}{c}
(1,1),(1,\delta)\\
(1,1),(1,\gamma) 
\end{array}
\right]\\
\fl\qquad\qquad\qquad\qquad\times
\frac{1}{\beta \tau}H_{1,1}^{1,0}\left[\frac{\mu_t^\beta\tau}{t^\beta\mu_\tau}\middle\vert
\begin{array}{c}
(0,\beta) \\
(0,1)
\end{array}
\right]
du\,d\tau,
\end{eqnarray}
which expresses in term of another Fox H function
 \begin{eqnarray}
\fl 
 S(t|x_0)
 =\frac{1}{\beta\sqrt{\pi}} H_{3,4}^{3,1}\left[ \left(\frac{x_0}{2\mu_x}\right)^{\alpha\gamma}\left(\frac{\mu_t}{t}\right)^{\beta\delta}\middle\vert
 \begin{array}{c}
  (1,1),(0,\beta\delta),(1,\delta)
 \\
 (0,\delta),(1,1),(1/2,\alpha\gamma/2),(0,\alpha\gamma/2)
 \end{array}
 \right].
 \end{eqnarray}

The asymptotic behavior of the Fox H function has been investigated in \cite{Mathai2010}. The probability density of the first-passage times is then obtained as $\rho(t|x_0) = - \partial S(t|x_0)/\partial t$.
\section{Conclusion}
In this article, we investigated the probability density $P(x,t)$ of finding a particle, started at time $0$ from the origin, at point $x$ at time $t$ subject to non-Gaussian diffusion of mixed origins.
We showed that various physical mechanisms that can be described by subordinated Brownian motion, can be combined and investigated as a whole. The necessary step is to calculate the probability density of the integrated diffusivity (or equivalently internal time). Therefore we proposed a model of continuous-time random integrated diffusivity for which the asymptotic behavior in two regimes described by an ordinary or fractional advection equation, is explicitly derived. From the combined subordination technique and the CTRID model, we derived the propagator $P(x,t)$ of the superposition of Gaussian processes with integrated diffusivity modeled as a CTRID as well as the space-diffusivity-time fractional model. We showed that combined heterogeneity delivers rich phenomenology; for example, the probability density of displacements can be non-Gaussian at short-time and converge to a second non-Gaussian distribution at long-times.
Finally we showed how the first-passage time statistics in any bounded domain in $\mathbb{R}^d$ (as well as in the half-line) can be obtained for the mixed models if the results are known for Brownian motion. The present work aims at shedding light on diffusion in disordered media in which several disorder mechanisms are simultaneously present.


\end{document}